\pdfoutput=1

\documentclass[11pt]{article}

\usepackage[]{acl}

\usepackage{times}
\usepackage{latexsym}
\usepackage{tabularx}
\usepackage{graphicx}
\usepackage{booktabs}
\usepackage{multirow}
\usepackage[ruled]{algorithm2e}
\usepackage{amsfonts}
\usepackage{amsmath}
\usepackage{footnote}
\usepackage{pifont}
\usepackage{lipsum}

\newcommand\blfootnote[1]{%
  \begingroup
  \renewcommand\thefootnote{}\footnote{#1}%
  \addtocounter{footnote}{-1}%
  \endgroup
}

\usepackage[T1]{fontenc}

\usepackage[utf8]{inputenc}

\usepackage{microtype}

\title{THE-X: Privacy-Preserving Transformer Inference with \\ Homomorphic Encryption}
\author{Tianyu Chen\textsuperscript{\ding{168} \ding{169} $\star$}, Hangbo Bao\textsuperscript{\ding{170}}, Shaohan Huang\textsuperscript{\ding{170}}, Li Dong\textsuperscript{\ding{170}}, Binxing Jiao\textsuperscript{\ding{171}}, \\ \textbf{Daxin Jiang\textsuperscript{\ding{171}}, Haoyi Zhou\textsuperscript{\ding{168} \ding{169}}, Jianxin Li\textsuperscript{\ding{168} \ding{169} \ding{41}}, Furu Wei\textsuperscript{\ding{170}}} \\
       BDBC, Beihang University, China\textsuperscript{\ding{168}} \\
       SKLSDE Lab, Beihang University, China\textsuperscript{\ding{169}} \\
        Microsoft Research\textsuperscript{\ding{170}} \ \  NLP Group, Microsoft STCA\textsuperscript{\ding{171}} \\
        \{tianyuc, zhouhy,lijx\}@buaa.edu.cn  \\ 
        \{t-habao,shaohanh, lidong1, binxjia, djiang, fuwei\}@microsoft.com
}

\begin{document}
\maketitle
\begin{abstract}


As more and more pre-trained language models adopt on-cloud deployment, the privacy issues grow quickly, mainly for the exposure of plain-text user data (e.g., search history, medical record, bank account). Privacy-preserving inference of transformer models is on the demand of cloud service users. To protect privacy, it is an attractive choice to compute only with ciphertext in homomorphic encryption (HE). However, enabling pre-trained models inference on ciphertext data is difficult due to the complex computations in transformer blocks, which are not supported by current HE tools yet. In this work, we introduce \textit{THE-X}, an approximation approach for transformers, which enables privacy-preserving inference of pre-trained models developed by popular frameworks. \textit{THE-X} proposes a workflow to deal with complex computation in transformer networks, including all the non-polynomial functions like GELU, softmax, and LayerNorm. Experiments reveal our proposed \textit{THE-X} can enable transformer inference on encrypted data for different downstream tasks, all with negligible performance drop but enjoying the theory-guaranteed privacy-preserving advantage.\blfootnote{\textsuperscript{${\star}$} Contribution during internship at MSRA.
\textsuperscript{\ding{41}}The corresponding author is Jianxin Li  <lijx@buaa.edu.cn>.}


\end{abstract}

\section{Introduction}

Accompanying the revolution of pre-trained models in many NLP applications, such as sentiment analysis~\cite{xu2019bert}, question answering~\cite{yang2019end}, information retrieval~\cite{yang2019simple}, and text generation~\cite{raffel2019exploring}, many related technologies have been deployed on the cloud to process user data from personal customers, small businesses, and large enterprises by industrial service providers. However, the convenience of the on-cloud pre-training technology also comes with a series of privacy challenges due to the sensitive nature of user data. For example, the input text or even text vector representations in user requests can leak private information, which may cause the specific user to be identified~\cite{schwartz2011pii, zhu2020deep}. This lack of privacy guarantees may impede privacy-conscious users from releasing their data to service providers. Thus, service providers may suffer from the deficiency of evolving models with user data. Besides, unintended data disclosure and other privacy breaches may result in litigation, fines, and reputation damages for service providers. These concerns spark our proposal of \textit{THE-X}, to enable privacy-preserving inference of transformer.

\begin{figure}
    \centering
    \includegraphics[width=\linewidth]{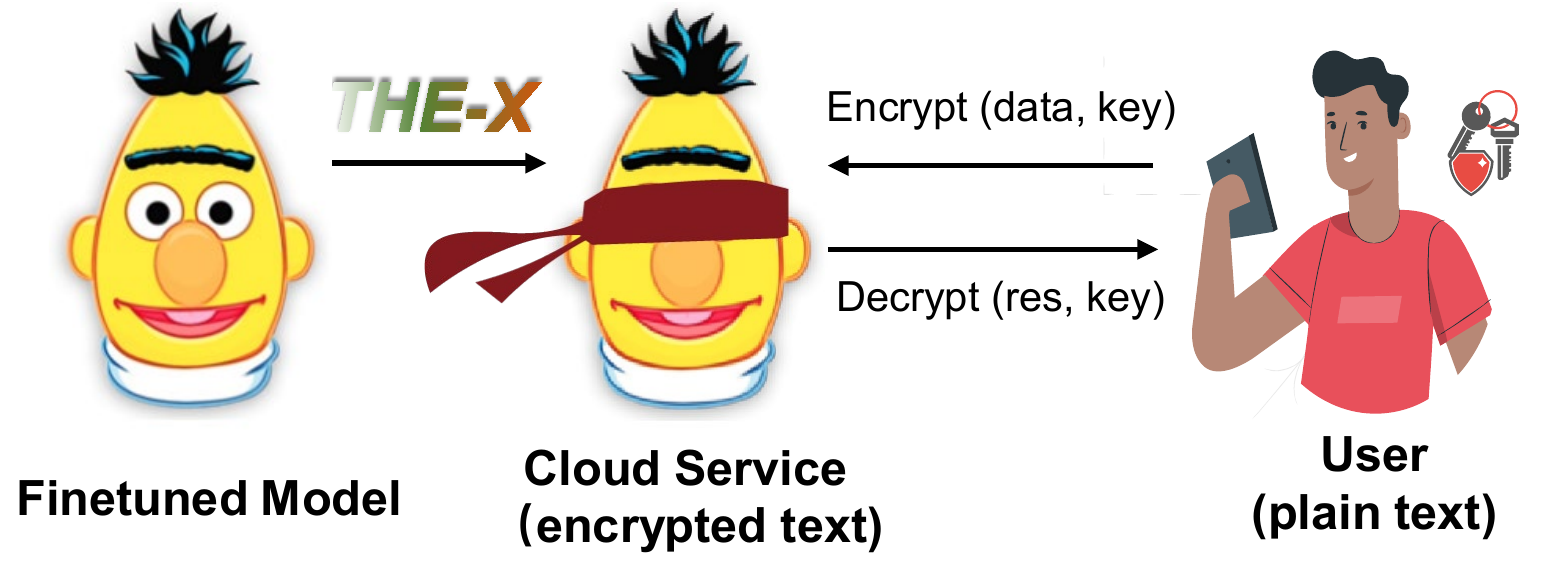}
    \caption{An overview of our \textit{THE-X}. The transformer-based model could inference on encrypted data with our \textit{THE-X}, enabling theory-guaranteed privacy protection for users. }
    \label{fig:overview}
\end{figure}

Specifically, we identify two challenges for the privacy-preserving inference of pre-trained models. The first challenge is how to protect users' plain text data from access by third-party service providers. (e.g., the clinic record or shopping history). Prior work has applied Differential Privacy (DP)~\cite{dwork2006calibrating} and its variants to address similar privatization issues - originally for statistical databases and more recently for DL~\cite{abadi2016deep} and NLP~\cite{qu2021natural, basu2021benchmarking, fernandes2019generalised, lyu2020differentially, basu2021privacy}. However, this solution may suffer from eavesdropping attackers. A handful of research~\cite{zhu2020deep, zhao2020idlg} demonstrated it possible to recover raw data from gradient leakage. Also, privacy protection could never be theory-guaranteed. The second challenge is the performance concern, recent works like \textit{TextHide}~\cite{huang2020texthide} and \textit{FedNLP}~\cite{lin2021fednlp} leverages the federated learning~\cite{yang2019federated} to train model on encrypted data, at cost of considerable performance dropping. Focusing on the privacy of training data, they have not fully explored privacy-preserving inference.

To solve the concerns above, we depict one practice of privacy-preserving inference in Figure \ref{fig:overview}, where a fine-tuned language model could be converted into the cloud service mode with \textit{THE-X}, and process users' data with its eyes blind. During inference, the content of the user query is anonymous to the transformer model. The results of computation are also ciphertext, which only can be decrypted by the user's private key. 

In addition, we need a theory-guaranteed encryption solution like the homomorphic encryption (HE)~\cite{gentry2009fully} to convince both service providers and users of the privacy security in production scenarios. The semantic security of HE is guaranteed by lattice-based cryptography, and the HE computation results on ciphertext could be decrypted to the same results in plaintext, preventing performance reduction cost. The basic idea of homomorphic encryption is to perform computations on encrypted data without first decrypting it, which could fully ensure privacy in cloud-serving scenarios. It allows user data to be encrypted and out-sourced to commercial cloud environments for processing.


However, due to the complex operations (e.g., GELU activation) in transformer-based models, the popular partially homomorphic encryption solution, which only supports addition or multiplication, can not easily be adapted into scenarios of pre-trained models. Based on HE transformer backend~\cite{boemer2019ngraph1, boemer2019ngraph2, boemer2020mp2ml}, we designed a series of approximation components to fulfill the whole inference pipeline of the mainstream transformer backbone.  We evaluate \textit{THE-X} for BERT-tiny on the GLUE benchmark~\cite{wang2018glue} and the CONLL2003 task~\cite{sang2003introduction}. Our results show that \textit{THE-X} can achieve the privacy-preserving inference with the averaged performance reduction of only 1.49\%.

Our contributions include:
\begin{itemize}
    \item We are the first work to explore the privacy-preserving transformer inference with HE.
    \item We design a practical and effective approximation workflow for converting transformer-based models into a function that consists of fully HE operations.
    \item A thorough set of experiments confirms the negligible performance reduction with our proposed \textit{THE-X} approximation.
\end{itemize}

\section{Background}

\subsection{Security and Privacy Concern of Pre-trained Models}

Pre-trained models like BERT~\cite{devlin2018bert} and GPT-3~\cite{brown2020language} rely heavily on the use of plain text data to get human-like performance. Despite the remarkable achievements of pre-trained models, these state-of-the-art models can not directly answer some sensitive use cases, including the medical record~\cite{christoph2015secure}, search history~\cite{shen2007privacy} and other personally identifiable information (PII). 

To avoid the direct computation on plain-text data, recent works like \textit{TextHide}~\cite{huang2020texthide} and \textit{DP-finetuning} ~\cite{kerrigan2020differentially} introduce the classical federated learning and differential privacy (DP) to protect the sensitive data. However, \textit{TextHide}~\cite{huang2020texthide} can only be applied to sentence-level tasks. Due to the mix-up operation, \textit{TextHide} fails to model token-level tasks like named entity recognition or semantic role labelling. \textit{DP-finetuning} would greatly sacrifice the performance of fine-tuned model by 20\% perplexity for a generation model like GPT-2.

\subsection{Practical Homomorphic Encryption}

The classic definition of homomorphic encryption is a form of encryption that permits users to perform computations on its encrypted data without first decrypting it. These computations results are retained in an encrypted form, which could be decrypted into identical output produced by the same computations on the unencrypted data. Let $F$ be a function  or the entire pre-trained model, $E$ as an encryption function, $D$ as a decryption function. Then for any allowed plain text input $x$, we have:

\begin{equation}
    F(x) = D(g(E(x)),
\end{equation}
where $g$ is a constructed function to play the same role of function $F$, except on encrypted data. Figure \ref{fig:overview} shows how a user performs inference using a cloud-deployed pre-trained model which is not trusted. First, the pre-trained model receives a ciphertext encrypted by the user private key and performs inference function $g$ on the ciphertext. Then, the server will send an encrypted result to the user, which can only be decrypted by the user key. At no point does the cloud service provider gain access to the plain text.

\begin{figure*}
    \centering
    \includegraphics[width=\linewidth]{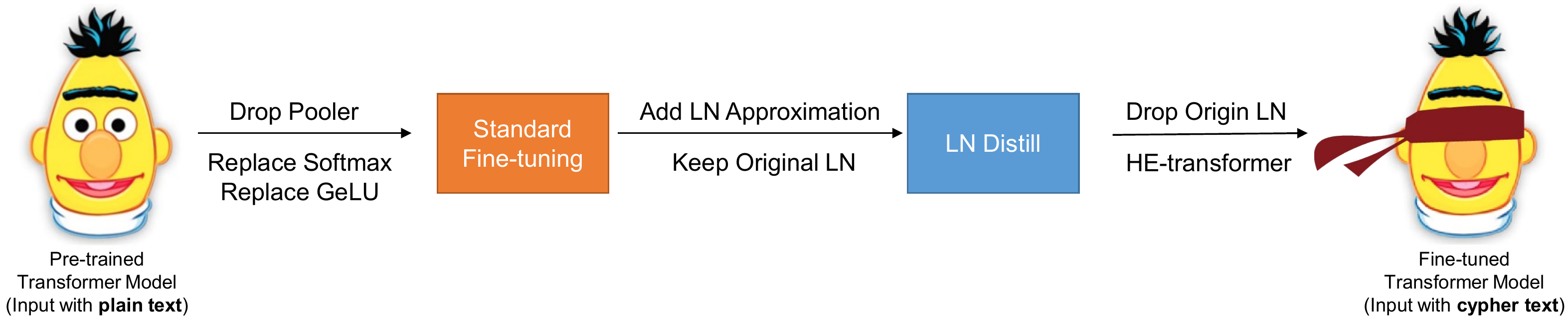}
    \caption{The Approximation Workflow of \textit{THE-X}. To replace the non-polynomial operations, we split the fine-tuning stage into several subphases. Given a pre-trained checkpoint, we drop the pooler of the pre-trained model and replace softmax and GeLU. Afterward, we follow the standard fine-tuning for classification or regression tasks. We add LayerNorm approximation into the fine-tuned model and distill knowledge from original LN layers. After dropping the original LN, we convert the model into fully HE-supported ops with the HE transformer.}
    \label{fig:workflow}
\end{figure*}

The Intel HE transformer for nGraph~\cite{boemer2019ngraph1, boemer2019ngraph2} is a Homomorphic Encryption backend to the deep learning models. Currently, it supports the CKKS~\cite{cheon2017homomorphic} encryption scheme, implemented by the Simple Encrypted Arithmetic Library (SEAL)~\cite{sealcrypto} from Microsoft Research. It is a research tool to demonstrate the feasibility of HE on deep learning.

\subsection{Challenges of Transformer Inference with HE}

Some HE schemes only support a single algebraic operation, such as addition or multiplication. These are known as "partially homomorphic" schemes (PHE). Other schemes, called "fully homomorphic"(FHE), support two such as addition and multiplication. Note that composing addition and multiplication suffices to construct polynomial functions, and hence polynomial approximations to non-polynomial functions such as GELU~\cite{hendrycks2016gaussian} or LayerNorm~\cite{xu2019understanding}. Notably, this limitation prevents the exact computation of any comparison-based operations such as Max, Min, as well as common functions such as exponential or sigmoid. Finally, "leveled homomorphic" schemes (LHE) support addition and multiplication, only up to a fixed computational depth.

\section{\textit{THE-X}: Formal Description}

There are two core ideas in \textit{THE-X}. The first one is to incorporate the user device into the HE inference, and the second is using "simplified computation" to approximate the non-polynomial functions.

In the following, we will describe how to enable homomorphic encryption of transformer-based models with \textit{THE-X}.

\subsection{Approximation Workflow}

\setlength{\algomargin}{1em}
\begin{algorithm}
\caption{Approximation Workflow}
\label{alg:app}
\DontPrintSemicolon
\SetAlgoLined
\SetKwFor{ForAll}{for all}{do}{end}
\KwData{labeled task data $\mathcal{D}$.}
\KwIn{pre-trained Transformer model $\mathcal{M}$, softmax estimation model $\mathcal{S}$.}

\nl $\widehat{\mathcal{M}} \leftarrow \mathcal{M} \odot (\mathcal{S}, ReLU) $. \\
\tcp*{\small{replace GELU and Softmax}}

\nl \While{not done}{
\nl    sample batches $(x_i, y_i)$ from $\mathcal{D}$, \\
\nl    let $(x_i, y_i)$ optimize $\widehat{\mathcal{M}}$ with $\mathcal{S}$ frozen. \\
}

\nl $\tilde{\mathcal{M}} \leftarrow \widehat{\mathcal{M}} \oplus \tilde{\mathcal{N}}$. \\
\tcp*{\small{add the layernorm approximation}}

\nl \While{not done}{
\nl    sample batches $(x_i, y_i)$ from $\mathcal{D}$, \\
\nl    freeze the parameters of  $\tilde{\mathcal{M}}$ except $\tilde{\mathcal{N}}$. \\
\nl    compute $k$-th layernorm output $O_k$, $\tilde{O_k}$. \\
\nl    compute loss $\ell_k$ = MSELoss($O_k$, $\tilde{O_k}$). \\
\nl    update $\tilde{\mathcal{N}}$ with loss $\mathcal{L} = \sum_k{\ell_k}$.
}

\nl $\Bar{\mathcal{M}} \leftarrow \tilde{\mathcal{M}} \ominus \mathcal{N}$. \\ \tcp*{\small{discard the origin layernorm}}
\KwRet $\Bar{\mathcal{M}}$.
\end{algorithm}

First, we present the approximation workflow of \textit{THE-X}, which consists of two stages: Standard Finetuning and LN Distill as depicted in Figure \ref{fig:workflow}. Given a pre-trained model $\mathcal{M}$ and corresponding downstream data, we aim to produce a fully HE supported $\Bar{\mathcal{M}}$ which is fine-tuned and ready for deployment.

The two-stage optimization of algorithm \ref{alg:app} aims to find the best approximation checkpoint. For computation efficiency,  pre-trained models can also be fine-tuned together with the layernorm approximation, and it needs only a single optimization loop. We will discuss the schedule of the different approximation workflow in Sec \ref{sec:schedule}. There are three major non-polynomial functions in the transformer block, where we will study in detail.

\subsubsection{Gaussion Error Linear Units (GLEU)}

With a computation of Gaussian error, Gaussian Error Linear Units (GLEUs)~\cite{hendrycks2016gaussian} is not suitable to serve as an active function in HE state. The Gaussian kernel includes unsupported functions like exponential. While in the implementation of the transformer, GELU is defined as a fast approximated version, where the $tanh$ function is still non-polynomial, unsupported by HE.

\begin{equation}
    G(x) = 0.5x(1 + tanh[\sqrt{2/\pi}(x + 0.044715x^3)]).  
\end{equation}

\begin{figure}
    \centering
    \includegraphics[width=\linewidth]{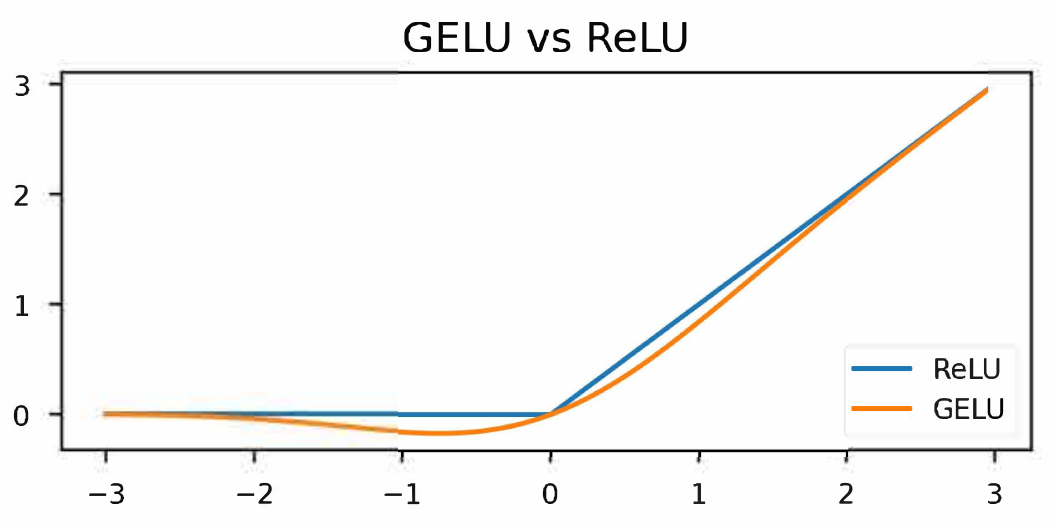}
    \caption{The activation results of GELU compared with ReLU. With an input around zero, the activation results are very close. With a larger or smaller input value, the activation results tend to converge.}
    \label{fig:GELU}
\end{figure}

We illustrate the numerical comparison between GELU and RELU in Figure \ref{fig:GELU} , where the outputs of GELU are very close to RELU. Hence, we propose to replace the GELU layer in the model with a ReLU activation function. Despite the $Max$ function in ReLU, other computations are well supported by HE. To enable the computation of $Max$, we implement the first key idea, incorporating the user device into the inference. The server will convey ciphertext input to the user for local $Max$ computation. Once received the connection, a user device decrypts the ciphertext input and calls the local $Max$ function to get the results and return re-encrypted results to the server. Despite the communication cost, no plaintext is leaked during the TLS connection and semantic security is guaranteed.

\subsubsection{Softmax}

The second non-polynomial function is softmax, which includes the exponential and division computation.

\begin{equation}
    Softmax(x_i) = \frac{exp(x_i)}{\sum_j exp (x_j)}.
\end{equation}

The first thought to approximate softmax is to find alternatives of softmax operation in transformer, which include Taylor series approximation~\cite{vincent2014efficient}, softmax-free linear attention~\cite{lu2021soft}. However, both of them have some limitations. The Taylor series approximation can only approximate the exponential operation. Softmax-free linear attention utilizes newton-inverse to approximate division, but the approximation error is unbounded in full-scale attention settings.

For these considerations, we have no choice but to design an estimation network with addition and multiplication. 

\begin{equation}
\label{eq:softmax_agent}
    S(x_i) = x_i \ast T( \sum_j ReLU(((x_j)/2 + 1)^3) ).
\end{equation}

Equation \ref{eq:softmax_agent} is the formal description of our softmax estimation network. Same as the approximation of GELU, ReLU operation here is realized by communication with the client. Instead of a division operation, we approximate reciprocal operation with a three-layer linear neural network denoted as $T$. 

To get a better estimation of softmax, we randomly generate input tensors whose values are between $[-3, 3]$ and use their softmax scores as MSE targets. Then we optimize the $T$ for 100k steps with a learning rate of 1e-3 until the MSE loss drop down to 1e-6.

An under-explored problem here is the \textit{Infinite value of Masked Attention}, where the input of softmax is always the masked attention scores.  To prevent the attention of masked tokens, the origin transformer model fills the masked attention scores with negative infinity before softmax. When fed with an infinite value, the softmax estimation model may face numerical disaster. We will discuss this phenomenon and the corresponding solution in Sec \ref{sec:attention overflow}.


\subsubsection{LayerNorm}
Recall that the layer normalization~\cite{ba2016layer} in transformer is implemented over a mini-batch of inputs, which could be formulated as:
\begin{equation}
    y = \frac{x - E[x]}{\sqrt{Var[x] + \epsilon}} \ast \gamma + \beta.
\end{equation}

The mean and standard deviation are calculated over division operations where the approximation is needed. $\gamma$ and $\beta$ are learnable affine transform parameters. To avoid the introduction of new parameters, we keep the learnable parameters while leaving the mean and standard deviation achieved by regression.

\begin{equation}
   \hat{y} = x \circ \boldsymbol{\gamma} + \beta.
\end{equation}

The new parameter $\boldsymbol{\gamma}$ predicts the value of standard deviation by regression from origin $\hat{\gamma}$. We find the simple linear replacement is enough for values with a small scale of bias. Here $\boldsymbol{\gamma}, \beta \in \mathbb{R}$ and $\circ$ denotes the Hadamard product.

The layer normalization will be applied in each multi-head attention block and after the output dense layer. So the approximation error tends to accumulate when the transformer stacks with too many layers. 

We treat the layernorm approximation as an individual stage in Figure \ref{fig:overview} as \textit{LN-Distill} to learn from origin LN layers. A challenge here is the \textit{Attention Overflow}, where the input attention score before normalization may have an unbounded scale, leading to numerical problems. We will discuss the detail of \textit{Attention Overflow} in Sec \ref{sec:attention overflow}.  

\subsubsection{Other Practical Replacement}
After the approximation workflow, a fine-tuned model consists of only addition and multiplication operations, which is fully compatible with homomorphic encryption. We power the model by HE transformer backend. Since the HE transformer backend could only work for TensorFlow checkpoint, any pre-trained transformers inherited from PyTorch building version need to be converted into TensorFlow format first. There are some other details worth mentioning here.

\begin{itemize}
    \item For the $softmax(\frac{QK^T}{\sqrt{d_k}}) V$ operation in attention score computation, we absorb the value of $\frac{1}{\sqrt{d_k}}$ into the weights of query projection layer.
    \item We use a fully kernel convolution layer instead of linear projection due to the lack of  supported dense operation.
    \item All matrix multiplication will be converted into the element-wise style.
    \item We drop the pooler layer for the unsupported operation of tanh.
\end{itemize}

\subsection{Privacy-preserving Inference}

In this section, we describe the behavior of HE models during privacy-preserving inference. Note that inference is completed by the joint effort of the server and the user device.

\begin{algorithm}
\caption{Inference with HE}
\label{alg:infer}
\DontPrintSemicolon
\SetAlgoLined
\SetKwFor{ForAll}{for all}{do}{end}
\KwIn{user plain text query $\mathcal{P}_q$, private key $\mathcal{K}$ generated under server protocol, encrypted server model $\mathcal{M}$.  }
\nl client computes embeddings: $\mathcal{E}_q \leftarrow \mathcal{P}_q $. \\

\nl client encrypts query embeddings: $\mathcal{C}_q \leftarrow Encrypt(\mathcal{E}_q, \mathcal{K})$. \\

\nl server forwards the model: $\mathcal{C}_i$ = $\mathcal{M}(\mathcal{C}_q)$. \\

\nl client handles activation: $\mathcal{C}_a$ = $ReLU(\mathcal{C}_i)$. \\

\nl server continues forwarding: $\mathcal{C}_o = \mathcal{M}(\mathcal{C}_a)$. \\

\nl client decrypts results: $\mathcal{P}_o = Decrypt(\mathcal{C}_o, \mathcal{K})$. \\

\end{algorithm}

In Algorithm \ref{alg:infer}, notably absent is the support of ReLU operations, where the server exchanges the activation results with the client. However, all the communication between client and server is in ciphertext, ensuring the privacy of user queries and may prevent eavesdropping attackers from recovering private text data.

\section{Experiments}

In this section, we design both sequence-level and token-level tasks to evaluate the approximation performance of our \textit{THE-X} solution. We also discuss several identified factors which greatly affect approximation workflow. 

\begin{table*}[htbp]
  \centering
  \caption{Performance on the GLUE\footnotemark[1] tasks for both baseline (standard finetuning) and \textit{THE-X} with BERT-tiny, measured on the development sets. We report the best results by hyper-parameter search. $|\mathcal{D}|$ denotes the number of training examples. \textit{THE-X} only suffers average utility performance loss: < 1.5\% in most tasks. `P/S corr.' is Pearson/Spearman correlation and `m/mm' denotes the accuracy scores on matched/mismatched set.}
  \resizebox{\linewidth}{!}{
    \begin{tabular}{cccccccccc}
    \toprule
    Tasks & $|\mathcal{D}|$  & Type  & Metrics & Baseline  & ReLU  & ReLU-S & ReLU-S-L & HE  & Perf $\downarrow$  \\
    \midrule
    SST-2 & 67k   & Sentiment & Acc.   & 82.45  & 82.40  & 82.34  & 82.11  & 82.11  & 0.34  \\
    MRPC  & 3.7k  & Paraphrase & F1/Acc. & 81.57/70.10 & 81.69/70.34 & 80.81/69.85 & 79.93/68.87 & 79.94/68.87 & 1.63/1.23 \\
    STS-B & 7k    & Similarity & P/S corr. & 72.83/73.66 & 72.89/73.03 & 74.19/74.27 & 68.38/70.96 & 68.39/70.97 & 4.44/2.69 \\
    QQP   & 364k  & Paraphrase & F1    & 80.28/84.03 & 79.55/82.89 & 79.38/83.36 & 78.28/83.75 & 78.33/83.63 & 1.95/0.40 \\
    MNLI  & 393k  & NLI   & m/mm & 69.75/70.75 & 69.51/70.60 & 68.61/69.13 & 68.59/69.41 & 68.47/69.08 & 1.28/1.67 \\
    QNLI  & 108k  & NLI   & ACC.   & 78.38  & 78.35  & 78.33  & 78.33  & 78.20  & 0.18  \\
    RTE   & 2.5k  & NLI   & ACC.   & 58.56  & 58.32  & 58.27  & 58.12  & 58.12  & 0.44  \\
    \midrule
    \multicolumn{4}{c}{Average Perf $\downarrow$} & 0.00     & 0.25  & 0.34  & 1.42  & 1.48  & \textbf{1.48} \\
    \bottomrule
    \end{tabular}}
  \label{tab:main_result}%
  
\end{table*}

\begin{table}[htbp]
  \centering
  \caption{Performance on the CONLL2003 task for both baseline and \textit{THE-X} with BERT-tiny, measured on the development sets. We find that the replacement with ReLU has a slight effect on performance and even gets a better F1 score by 0.12 than original GELU activation.}
  \resizebox{\linewidth}{!}{
    \begin{tabular}{ccccc}
    \toprule
    Metrics & Precision & Recall & F1  & Perf $\downarrow$ \\
    \midrule
    Raw   & 82.34 & 84.85 & 83.57 & 0 \\
    ReLU  & 82.29 & 85.13 & 83.69 & -0.12 \\
    ReLU-S & 82.08 & 84.73 & 83.38 & 0.19 \\
    ReLU-S-L & 79.65 & 83.79 & 81.67 & 1.90 \\
    HE    & 79.65 & 83.79 & 81.67 & \textbf{1.90} \\
    \bottomrule
    \end{tabular}
    }
  \label{tab:NER_results}
\end{table}

\subsection{Evaluation Tasks}

GLUE~\cite{wang2018glue}, the General Language Understanding Evaluation benchmark, is a collection of tools for evaluating the performance of models across a diverse set of existing NLU tasks. We choose a subset of GLUE\footnote[1]{CoLA task is not reported because of the limited capacity of BERT-tiny.} tasks, which include: MRPC~\cite{dolan2005automatically}, SST-2~\cite{socher2013recursive}, QQP\footnote[2]{https://www.quora.com/profile/Ricky-Riche-2/First-Quora-Dataset-Release-Question-Pairs}, STS-B~\cite{cer2017semeval}, MNLI~\cite{williams2017broad}, QNLI~\cite{rajpurkar2016squad}, and RTE~\cite{dagan2005pascal, haim2006second, giampiccolo2007third, bentivogli2009fifth}.


Following previous work~\cite{devlin2018bert, turc2019well}, we exclude the WNLI task from the GLUE benchmark. We also use the famous CoNLL-2003~\cite{sang2003introduction} named entity recognition task as our additional token-level evaluation. In conclusion, we include the most varieties of NLU tasks, covering both sequence-level and sentence-level tasks, in both regression and classification format.

\subsection{Experiment Settings}

For computation efficiency and energy-saving consideration, we use the released BERT-tiny~\cite{turc2019well} as our demo model, which is a standard transformer-based language model with only 2 layers and a hidden size of 128. We provide four settings to evaluate different parts of our approximation components.

\begin{itemize}
    \item \textbf{Baseline.} In this setting, we make no replacement or approximation. We use the raw pre-trained checkpoint to fine-tune on downstream tasks.
    \item \textbf{ReLU.}  We fine-tune the pre-trained model with all GELU activation replaced with ReLU.
    \item \textbf{ReLU-S.} In addition to ReLU, we fine-tune the model with the softmax operation replaced by the softmax estimation model.
    \item \textbf{ReLU-S-L.} We implement full approximation including a layer normalization replacement.
    \item \textbf{HE.} We convert the fine-tuned checkpoint with HE-transformer and power the inference with SEAL backend.
\end{itemize}

\textbf{Implementation.} To reduce the variance of results under different settings, we choose hyper-parameters from a fixed set during approximation fine-tuning and HE inference runtime.

\begin{itemize}
    \item For fine-tuning the approximation components, we choose a batch size from \{4, 8, 16, 32, 128\} and a learning rate from {1e-4, 3e-4, 3e-5, 5e-5} as mentioned in the initial bert code~\cite{turc2019well}. We use an Adam optimizer with weight decay chosen from \{0.05, 0.1,  0.2, 0.4, 0.5\} 
    \item For HE evaluation, we use the HE-transformer backend, where two parameters are recommended searching by Intel, the poly modules and coeff-modules. We choose the poly modules degree from \{1024, 2048, 4096, 8192, 16384\} and choose the coeff-modules from \{20, 30, 60\}.
\end{itemize}

\subsection{Approximation Results}

Table \ref{tab:main_result} shows the results of the baseline and \textit{THE-X} on the GLUE benchmark. The averaged performance reduction of \textit{THE-X} is 1.48\% when compared to the baseline model. We observe the most performance reduction comes from the approximation of layernorm, which incurs a reduction of 1.08\%. The softmax estimation model contributes the least performance drop among the approximation components, for only 0.09\% on average, indicating the softmax function could be well imitated by neural networks. We also find the average performance reduction of HE is quite negligible, where the slight drop may be due to the sequence truncation.

The results of \textit{THE-X} on token-level NER task are reported in Table \ref{tab:NER_results}. The replacement of GELU with ReLU even improves the performance of the F1 score. We assume the slight improvement may come from unexpected bias. However, the layernorm approximation incurs the most performance reduction. We assume token-level tasks need a more detailed pattern in attention score. After all, \textit{THE-X} still works well in the token-level task with a merely F1 reduction of 1.9\%.

Across different types of tasks, we find our \textit{THE-X} yields the best performance on the classification tasks, including paraphrase, sentiment and NLI. Among the classification tasks, the performance of QNLI drops the least, for only 0.18\%. We also find the performance drops most on the regression tasks, such as the similarity task STS-B, for 4.44\% pearson correlation and 2.69\% spearman correlation. We assume the regression task needs a higher numerical precision than the classification task.


\begin{figure}
    \centering
    \includegraphics[width=\linewidth]{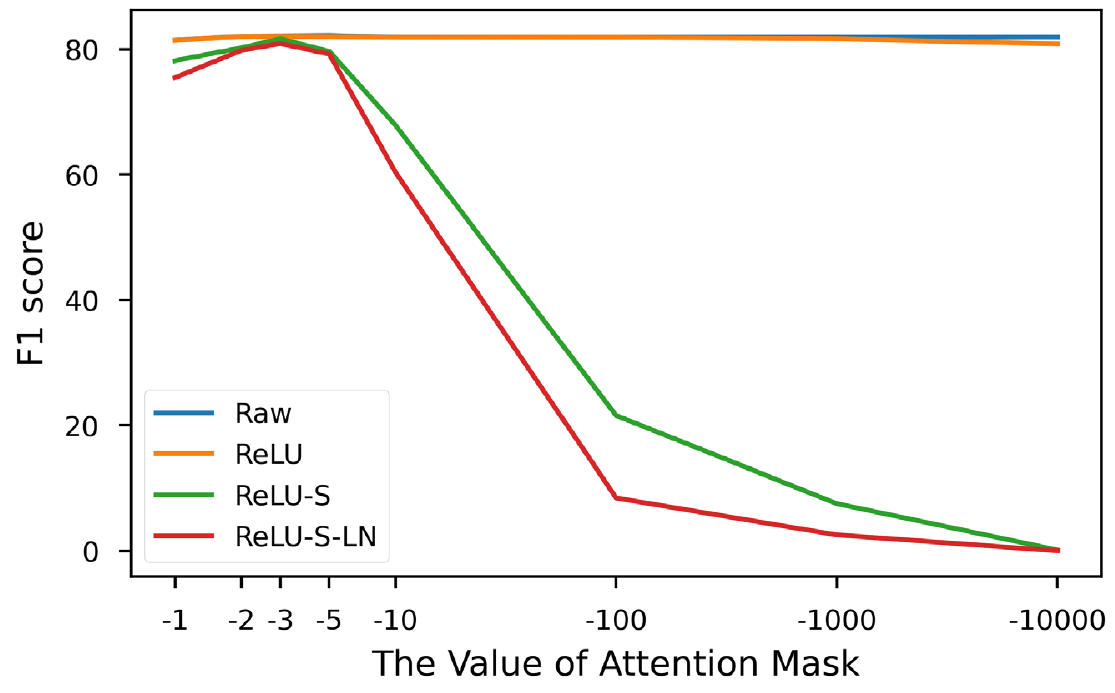}
    \caption{Performance on CONLL2003 task with different mask values. We find the “Negative Infinity” value of the 0 mask greatly reduces approximation performance. In \textit{THE-X} using a mask value in [-3, -5] might be a default choice.}
    \label{fig:mask_value}
\end{figure}

\begin{figure*}
    \centering
    \includegraphics[width=\linewidth]{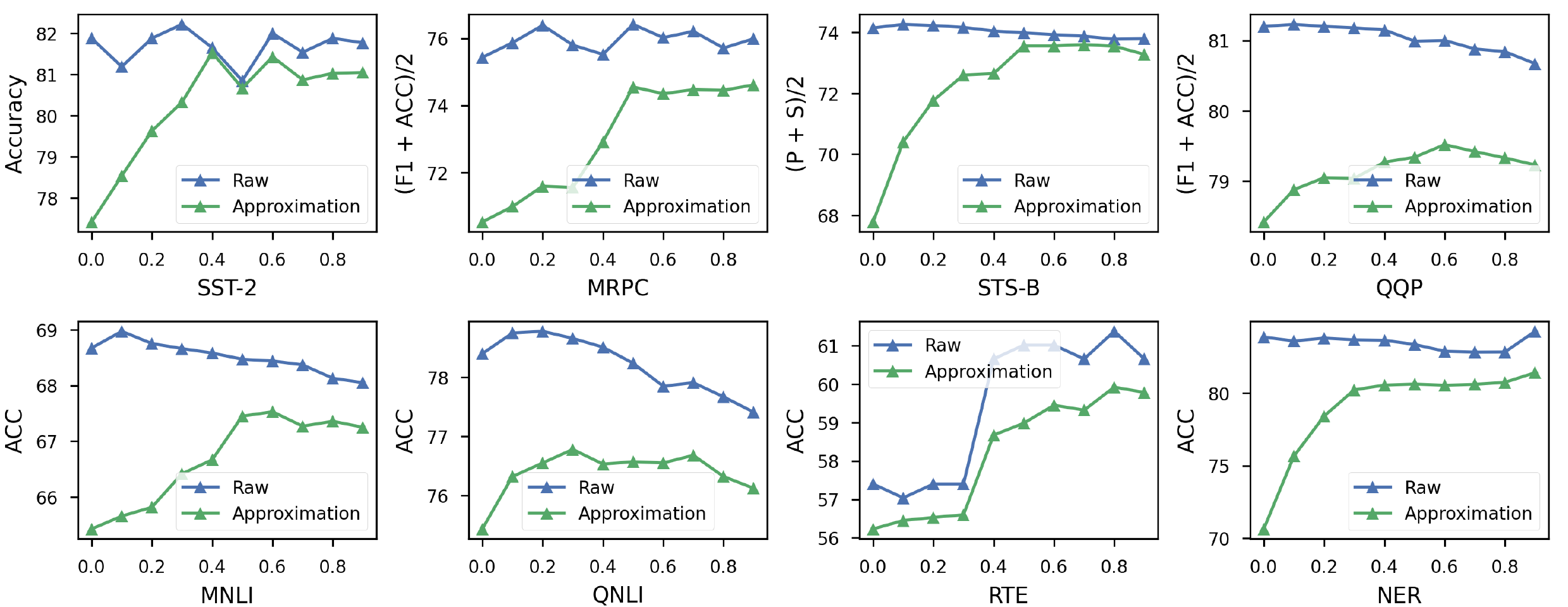}
    \caption{Performance on all tasks with different weight decay values, measured on the development sets. Metrics are marked on the y-axis and weight decay values are marked on the x-axis.}
    \label{fig:weight_decay}
\end{figure*}

\subsection{Negative Infinity}
\label{sec:infinite}

Recall in Equation \ref{eq:softmax_agent}, we replace softmax with a neural estimation model. To prevent the attention of masked tokens, the origin transformer model fills the masked attention scores with negative infinity before softmax, where the numerical disaster occurs in our approximation method. In Figure \ref{fig:mask_value}, to solve this problem, we give an empirical study of how "negative" the masked attention scores should be. Despite the indistinguishable F1 score change of raw model fine-tune with different attention mask values, the approximation method is extremely sensitive to the numerical changes. We assume the softmax estimation model fails to deal with large input values and leads to a credible performance drop. However, when the value of the attention mask becomes too small, it serves as a bias to attention scores, which also leads to a certain performance drop. We recommend using a moderate mask value between -2 and -5.

\subsection{Attention Overflow}
\label{sec:attention overflow}
Another challenge of \textit{THE-X} is the attention score input of layer normalization. In most cases, the scale of multi-head attention output is very dense around $[-1, 1]$. However, before normalization, we also observe the attention scores are scarily sparse, with some extreme values reaching $1e4$, which is difficult for our LN-distill stage. To prevent the overflow attention scores, we use the \textbf{weight decay} of Adam optimizer as regularization.

In Figure \ref{fig:weight_decay}, we present the attention overflow phenomenon across different tasks. Without any regularization, our approximation method yields uncontrolled attention scores, leading to poor performance. As the weight decay increases, the attention scores tend to converge and benefit better approximation results. We also observe that the larger weight decay may harm the performance on NLI tasks, where the regularization could be seen as trade-off between better approximation results and higher performance upper bound. For the NER task, larger weight decay may even benefit the performance and also boost our approximation method.

\subsection{Schedule of Approximation workflow}
\label{sec:schedule}

There are still doubts about how to organize the several optimization steps for the best approximation performance. We investigate four schedule plans:

\begin{itemize}
    \item \textbf{Two Stages.} Where we freeze the softmax estimation model during standard fine-tuning. We select the best checkpoint to implement the second stage - distill the layer normalization network.
    \item \textbf{Joint FT S.} We optimize the softmax estimation model during standard fine-tuning and apply the LN-distillation after.
    \item \textbf{Joint FT LN.} We apply one-pass optimization with the softmax estimation model frozen but update the other parameters including layer normalization network. No further LN-distill will be implemented.
    \item \textbf{Joint FT S + LN.} A total one-pass optimization with all approximation parameters updated with the model together.
\end{itemize}

\begin{figure}
    \centering
    \includegraphics[width=\linewidth]{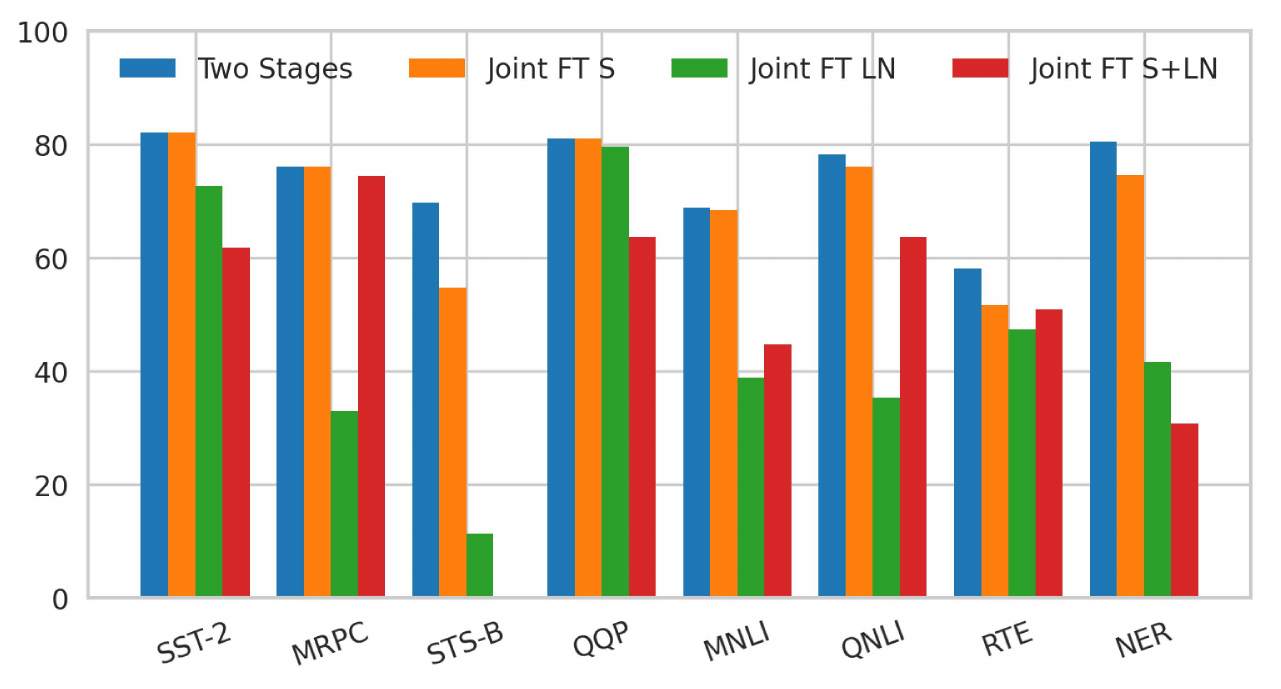}
    \caption{Performance on all tasks with different organizations of approximation workflow. Jointly fine-tuning the softmax estimation model or approximated layernorm leads to a performance drop across all tasks.}
    \label{fig:schedule}
\end{figure}

As illustrated in Figure \ref{fig:schedule}, we observe that fine-tuning the different approximation components individually (aka. "Two stages") may be a good default to keep the best performance of approximation. For the regression task STS-B, jointly fine-tuning the softmax estimation model and approximated layernorm even fails to fulfill the approximation pipeline, pulling the performance down to 0.4\%. We assume fine-tuning different components may fall into a \textit{bi-level} optimization problem and it is hard to achieve satisfying results. In conclusion, the softmax estimation model and the approximated layernorm are both critical components to the performance of \textit{THE-X}, deserving individual optimization.

\section{Conclusions}

We present \textit{THE-X}, a practical approach to enable pre-trained transformer models to infer under homomorphic encryption. It requires several approximation components to replace the original operations in the transformer model. It imposes a slight burden in terms of performance cost but enjoys the full advantage of homomorphic encryption - the theory-guaranteed user privacy.

We see this as a first step in combing homomorphic encryption to address emerging privacy issues in pre-trained models. We hope our work motivates further research, including better approximation solutions on different NLP applications. 

\section{Acknowledgments}
This paper is supported in part by the NSFC through grant No.U20B2053. We also thanks the support from Beijing Advanced Innovation Center for Future Blockchain and Privacy Computing.

\bibliography{anthology,custom}
\bibliographystyle{acl_natbib}




\end{document}